\documentclass[12pt]{spieman}  
\usepackage{amsmath,amsfonts,amssymb}
\usepackage{graphicx}
\usepackage{setspace}
\usepackage{tocloft}

\title{Status of the MEDUSAE post-processing method to detect circumstellar objects in high-contrast multispectral images.}

\author[a,*]{Faustine Cantalloube}
\author[b]{Marie Ygouf}
\author[c]{Laurent M. Mugnier}
\author[d]{David Mouillet}
\author[c]{Olivier Herscovici-Schiller}
\author[a]{Wolfgang Brandner}
\affil[a]{Max Planck Institut f\"ur Astronomie, K\"onigstuhl 17, D-69117, Heidelberg, Germany}
\affil[b]{ IPAC, Caltech, 1200 E. California Blvd., Pasadena, CA 9112, United States of America}
\affil[c]{Onera - Optics Department, 29 avenue de la Division Leclerc, F-92322 Chatillon Cedex, France}
\affil[d]{IPAG UJF-Grenoble 1 / CNRS-INSU, Institut de Planétologie et d’Astrophysique de Grenoble (IPAG) UMR 5274, Grenoble, F-38041, France}

\cftpagenumbersoff{figure}
\cftpagenumbersoff{table} 
\begin{document} 
\maketitle

\begin{abstract}
The MEDUSAE method (Multispectral Exoplanet Detection Using Simultaneous Aberration Estimation) is dedicated to the detection of exoplanets and disks features in multispectral high-contrast images. 
The concept of MEDUSAE is to retrieve both the speckle field and the object map via a stochastic approach to inverse problem (taking into account the statistics of the noise) in the Bayesian framework (using parametric regularization). 
One fundamental aspect of MEDUSAE is that the model of the coronagraphic PSF is analytic and parametrized by the optical path difference which, contrary to the phase, is achromatic. The speckle field is thus estimated by a phase retrieval, using the spectral diversity to disentangle the planetary signal from the residual starlight. 
The object map is restored via a non-myopic deconvolution under adequate regularization.
The basis of this MEDUSAE method have been previously published and validated on an inverse crime. In this communication, we present its application to realistic simulated data, in preparation for real data application. The solution we proposed to attempt bypassing the main differences between the model used for the inversion and the real data is not sufficient: it is now necessary to make the model of the coronagraphic PSF more realistic.
\end{abstract}

\keywords{Post-processing, high-contrast imaging, extreme Adaptive Optics, coronagraphy, exoplanet and disk imaging, AO4ELT Proceedings}

{\noindent \footnotesize\textbf{*}Faustine Cantalloube,  \linkable{cantalloube@mpia.de} }


\section{Introduction}
\label{sect:intro}  
Direct imaging of exoplanets provides three observables (projected separation from the host star, contrast to the host star and detection limit for the data set) that can be translated into astrophysical parameters (such as mass, radius, temperature or orbital parameters) via dynamical and evolutionary models. Statistical survey then complete the overview of the planet population as a function of the host star type and environment. The derived properties of the discovered systems are essential to discriminate between different planetary formation scenarios.

Within the direct imaging methods, high contrast imaging (HCI) gives access to a unique parameter space: for an 8-m class telescope working in the infrared wavelengths\footnote{The other direct imaging techniques being Very Long Baseline Interferometry (VLBI), reaching contrast down to $5.10^{-3}$ from about 100 mas and Sparse Aperture Masking (SAM), reaching contrast down to $5.10^{-3}$ from about 100 mas\cite{Cheetham2016}.}, we can currently reach a contrast down to $10^{-6}$ from about 300 mas\cite{Samland2017}, which corresponds to the typical giant exoplanets and brown dwarfs regime on wide orbits. This achievement is the result of a synergy between hardware solutions (combination of an extreme adaptive optics and coronagraph), observation strategies (observation and instrumental setup, specific calibration or turbulence profiling) and advanced image processing techniques.

In this communication, we focus on the image processing part, applied to multispectral data as provided by Integral Field Units (IFUs). In section 2, we describe the concept of the MEDUSAE method in the context of IFU data. In section 3, in order to show the typical performance of the method in an ideal context, MEDUSAE is applied on data simulated with the model used for the inversion (inverse crime). In section 4, we apply MEDUSAE on realistic data to infer the major problems to tackle before its application on on-sky data, which leads to our conclusion in section 5.

\section{Presentation of the MEDUSAE method}
\label{sect:medu1}
An essential ingredient of HCI is the post-processing techniques that are applied to the (cleaned, centered and rotated) images that allows scientists to reach $10^{-6}$ contrasts at 300 mas from the post-extreme adaptive optics (AO) coronagraphic images, in which the raw contrast is of typically $10^{-4}$. In the raw images, the contrast is limited by the starlight residuals showing speckles, arising from residual post-AO aberrations, hiding the faint circumstellar features. The key idea to build image processing for the post-AO coronagraphic images is to exploit a diversity in the image: a parameter along which the two components (circumstellar features and starlight residuals) show a different behavior. 
In the following we describe the specificity of multispectral data that are exploited by MEDUSAE. We then describe the working principle of MEDUSAE along with the main hypothesis on which it relies.

\subsection{Exploitation of the spectral redundancy for HCI}
Most of the latest and next generation of high-contrast instruments dedicated to exoplanet imaging are equipped with IFUs\cite{Antichi2008,Macintosh2008,Peters-Limbach2013} in order to increase the detection sensitivity, extract the spectra of planetary companions and constrain astrometric measurements thanks to the spectral redundancy.

\begin{figure}[!h]
\begin{center}
\begin{tabular}{c}
\includegraphics[height=3.5cm]{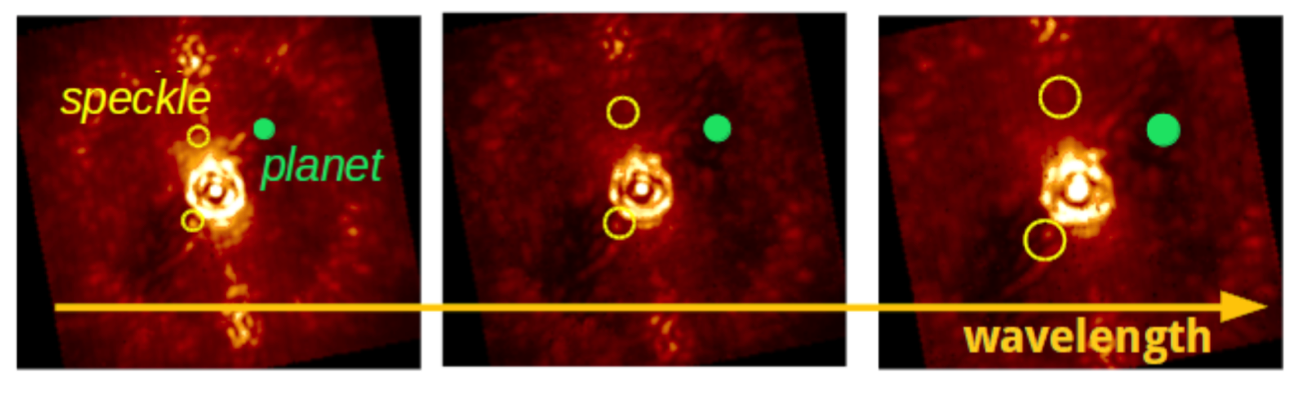}
\end{tabular}
\end{center}
\caption 
{ \label{fig:spectraldiv}
Images from the Spectro-Polarimeter High-contrast Exoplanet REsearch (SPHERE\cite{Beuzit2008}) Integral Field Spectrograph (IFS\cite{Antichi2008}) instrument installed at the Very Large Telescope (VLT, Chile), showing the spectral diversity: at larger wavelengths, the speckles are spreading radially in the field of view while planetary companions have a fix location.} 
\end{figure}

Our goal here is to disentangle the potential planetary signals from the starlight residuals in multispectral high contrast images. In such images, the starlight residuals are non-calibrated starlight leakages due to (i) the residuals from the adaptive optics correction and (ii) quasi-static optical aberrations within the instrument (change of temperature, pressure, gravitational bent etc.). In the focal plane image at the detector, (i) shows a smooth halo and (ii) is responsible for speckles of typical size $1\lambda/D$ ($\lambda$ being the wavelength and $D$ the pupil diameter) which are temporally evolving on scales from fraction of minutes to hours\cite{Hinkley2007}. 
Given the origin of each component, the spectral diversity brings the following: (1) the speckles are radially spreading in the field of view when the wavelength increases, whereas (2) the planetary signal's centroid remains fix (Fig.~\ref{fig:spectraldiv}).

All the post-processing techniques currently operational are empirical: they do not make use of any external knowledge about the instrument. A reference speckle field (also called coronagraphic PSF) is estimated directly from the data set and subtracted from the science images. Simultaneous Differential Imaging technique (SDI\cite{Racine1999,Sparks2002}) consists in rescaling the images at one common wavelength, to obtain the reference PSF which is subtracted from the science images. The companion(s) should ideally be absent from the reference image in order to avoid the self-subtraction of the planetary signal. However this rescaling is valid at first order and the presence of the coronagraph makes the actual relation highly non-linear (it is not a convolution process any longer), mostly at short separation. Moreover, this SDI, usually induces spectral correlation and contaminates the extracted photometry of the companion. On top of that, the signal noise ratio (S/N) of the companion and the contrast curve detection limits are, by definition, dependent upon the spectral type of the companion. At this stage, it is important to develop post-processing techniques avoiding the rescaling and subtraction. The MEDUSAE method intends to address these limitations by estimating the aberrations that are responsible for the speckle field instead of empirically assessing it from the image cubes. To do so, it uses a maximum a posteriori approach, described in the following.

\subsection{Principle of MEDUSAE}
\begin{figure}[!h]
\begin{center}
\begin{tabular}{c}
\includegraphics[height=4cm]{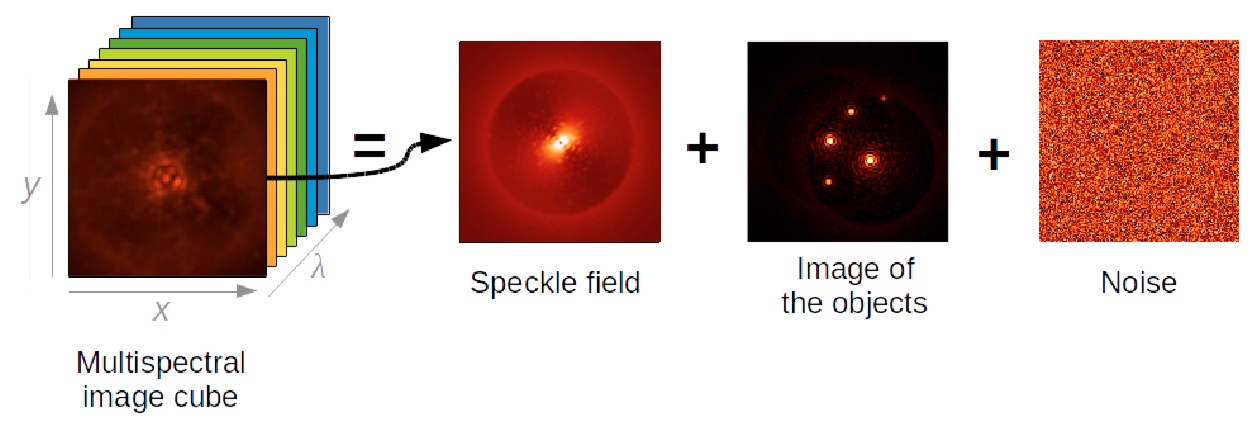}
\end{tabular}
\end{center}
\caption 
{ \label{fig:model}
Three components of the model of each image of the multispectral cube used in MEDUSAE.} 
\end{figure}

There is no operational post-processing method today which makes full use of the knowledge of the multispectral data production. The MEDUSAE solution is to extract information from the data by including the instrument knowledge and type of object sought. In MEDUSAE, we want to estimate both the speckle field and the planetary signals. Each image of the multispectral cube is modeled as the addition of (i) the speckle field, (ii) the object map and (iii) a noise map (see Fig.~\ref{fig:model}).

\begin{figure}[!h]
\begin{center}
\begin{tabular}{c}
\includegraphics[height=4.5cm]{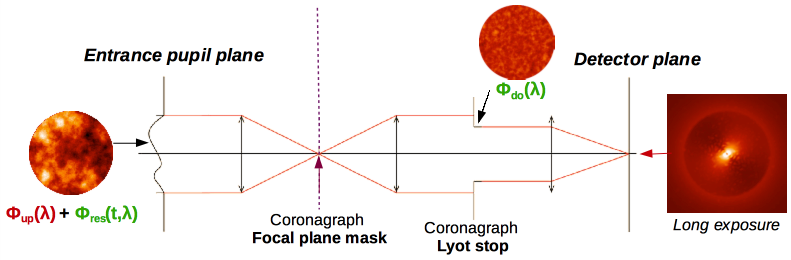}
\end{tabular}
\end{center}
\caption 
{ \label{fig:SFmodel}
Schematic view of the post-AO coronagraph set-up, from the entrance pupil (left) to the detector focal plane (right). The analytical long-exposure PSF expression is derived from a plane to plane light propagation of this set-up. The sources of aberration in green are assumed to be known while the red one is to be estimated.} 
\end{figure}

\textbf{Model of the speckle field:}
The speckle field is the convolution of the star (a Dirac peak of flux $f^{star}$) and a "coronagraphic PSF". As mentioned above, the model of coronagraphic PSF is parametrized by the aberrations in the system. More specifically we use the analytical model of long-exposure coronagraphic PSF published in Ref.~\citenum{Sauvage2010}. This analytical expression is a plane to plane propagation of the aberrations in the Fraunhofer diffraction framework. The aberrations, during one exposure time of several seconds, are separated into three terms, as in Fig.~\ref{fig:SFmodel}: (1) the very fast evolving aberrations originating from the AO residuals (on scale of fraction of seconds), (2) the temporally stable aberrations before the coronagraphic focal plane mask (called \emph{upstream aberrations}) and (3) the temporally stable aberrations after the coronagraphic focal plane mask (called \emph{downstream aberrations}). In the pupil plane, as the model is a long exposure, the AO residuals (1) are averaged into a phase structure function ($D_{\Phi}$), whereas the upstream (2) and downstream (3) aberrations are described via phase screens ($\Phi_{up}$ and $\Phi_{do}$ resp.). In the image plane, the AO residuals (1) translate into a smooth halo, whereas the upstream phase (2) is responsible for most speckles in the field of view and the downstream phase (3) has little impact on the final image. We thus assume that the phase structure function (1) and the downstream phase (3) are known via either dedicated calibration or external estimation (fit, filtering etc.): In MEDUSAE, only the upstream phase is estimated. As the phase depends on the wavelength, we instead estimate the optical path difference (opd) which is achromatic under the assumption that the aberrations are in pupil planes. In the end, only one opd screen estimation is necessary to describe the multispectral coronagraphic PSF. However, the coronagraphic PSF model is highly non-linear, which induces many degeneracies: one speckle field can be explained by various upstream opd maps.

In practice, to build the multispectral model from these aberrations, we numerically fixed the pupil diameter for all the wavelengths and we adapted the focal plane sampling by zero padding the pupil image. The resulting focal plane images are then cropped to the size of the data.\\

\textbf{Model of the object map:}
The object map, $o$, is simply the convolution between the object and the non-coronagraphic (off-axis) PSF. The non-coronagraphic PSF also depends on the three terms mentioned above, including the upstream optical path difference $\delta_{up}$ to be estimated.\\

\textbf{Model of the noise map:} 
This term, $n$, is constituted of the photon and the detector noises affecting the high contrast images. In the high flux regime, as it is the case here, the noise can be approximated by a Gaussian distributed noise whose variance $\sigma^2_n$ is given by the image\cite{Mugnier2004}.\\

The complete model $M$ of the multispectral high-contrats image cube then writes:
\begin{equation}
\label{eq:model}
M_{\alpha,\lambda}(o_{\alpha,\lambda}, \delta_{up}, f^{star}_{\lambda}) =
f^{star}_{\lambda} \times h^{coro}_{\alpha,\lambda}(\delta_{up}) + 
o_{\alpha,\lambda} \star  h^{non-coro}_{\alpha,\lambda}(\delta_{up}) +
n_{\alpha,\lambda} \, ,
\end{equation}
where $\alpha$ is the spatial dimension and $\lambda$ the spectral dimension. $h^{coro}$ and $h^{non-coro}$ are the long-exposure PSF, respectively coronagraphic and non-coronagraphic.\\

\textbf{Criterion to be minimized:}
Under the assumption of Gaussian noise, the corresponding maximum likelihood estimation is equivalent to a least-square minimization. We also add penalization to the resulting criterion, on the object (e. g. sparse spatial regularization for planets) and on the flux of the star (positivity constraint). Further regularization can be later included in that global criterion if we have a consistent knowledge about the unknowns (e. g. the root mean square of the upstream phase could be regularized). 
The criterion $J$ to be minimized with respect to our three unknowns, $o$, $f^{star}$ and $\delta_{up}$, then writes:
\begin{equation}
\label{eq:criterion}
J(o_{\alpha,\lambda}, \delta_{up}, f^{star}_{\lambda}) =
\sum_\lambda \sum_\alpha \frac{1}{2 \sigma^2_{n; \lambda, \alpha}} |i_{\alpha,\lambda} - 
[f^{star}_\lambda \times h^{coro}_{\alpha,\lambda}(\delta_{up}) + 
o_{\alpha,\lambda} \star  h^{non-coro}_{\alpha,\lambda}(\delta_{up}) ] |^2 + 
\mathcal{R}(o) + \mathcal{R}_\lambda(f^{star}) \, ,
\end{equation}
where $i_{\alpha,\lambda}$ is the multispectral images and $\mathcal{R}$ are the regularizations (here, on the object $o$ and on the star spectrum $f^{star}$).\\

\textbf{Minimization strategy:}
In order to minimize this criterion $J$, in a first step we assume the object is faint enough to ignore it and MEDUSAE estimates the upstream opd. Once a minimum has been reached, the estimated speckle field is assumed to be known and the object map is estimated. This alternated estimation of the speckle field and the object is repeated until the global criterion of Eq.~(\ref{eq:criterion}) reaches convergence\footnote{The convergence threshold (or stopping rule) is defined as $\epsilon = \frac{|J_i - J_{i-1}|}{|J_i + J_{i+1}|/2} $, where $J_i$ is the value of the criterion at the iteration number $i$.}. For more details about the numerical minimization in practice\cite{Ygouf2013,CantalloubePhD}, please refer to Sec.~\ref{sect:annexe}. 

Several relevant conclusions from the previously published paper\cite{Ygouf2013} on MEDUSAE are worth reminding here:
\begin{itemize}
\item Using distant spectral channels increases the sensitivity;
\item If the static (given) downstream aberrations are large, it helps disentangling the sources of the speckles and consequently the estimation of the upstream aberrations is better;
\item Because there is an inherent degeneracy between the star flux and the upstream aberrations to obtain the speckle field, if $D_\Phi$ is strong, it helps disentangling the star flux from the upstream aberrations which are consequently better estimated.
\end{itemize}

A recent publication presents a similar algorithm, PeX\cite{Devaney2017} (for Planet eXtractor), in which the speckle field model is a multispectral adaptation of the Taylor expansion describing a post-AO PSF\cite{Perrin2003} while the object is estimated in a clean-like approach.

\section{Application of MEDUSAE on ideal simulated data}
\label{sect:medu2}
In a first approach, we applied MEDUSAE to data that are simulated using the same model as the one used for the inversion. This so called \emph{inverse crime} test is useful to validate the method, to polish the minimization strategy and to perform sensitivity tests.

We simulated high-contrast images using the analytical long-exposure coronagraphic PSF model described in Sect.~\ref{sect:medu1}. 
The entrance pupil is defined as circular and uniform and the Lyot stop diameter is set equal to the entrance pupil diameter. 
The AO residual phase structure function $D_{\Phi}$ is simulated by compiling the error budget of an AO system\cite{Jolissaint2006}, whose parameters are defined in Tab.~\ref{tab:simuAO}. 
\begin{table}[ht]
\caption{Adaptive optics parameters used to simulate the AO residual phase structure function $D_{\Phi}$.} 
\label{tab:simuAO}
\begin{center}       
\begin{tabular}{|c|c|} 
\hline
Set-up & Value  \\
\hline\hline
Diameter of the telescope & 8 m  \\
\hline
Median seeing & $0.85$" at $0.5 \, \mu$m  \\
\hline
Number of turbulent layers & 3 \\
\hline
$C_n^2$ profile weight & $[40\% - 40\% - 20\% ]$ \\
\hline
Wind speed & $[12.5 - 12.5 - 12.5]$ m/s \\
\hline
Wavefront sensor & Shack-Hartmann working at 700 nm \\
\hline
Number of sub-pupils & $40 \times 40$  \\
\hline
Number of DM actuators & $41 \times 41$  \\
\hline
Loop frequency & 1.2 kHz  \\
\hline
Star magnitude & 8 at 700 nm  \\
\hline
Equivalent exposure time & 30 min  \\
\hline
\end{tabular}
\end{center}
\end{table} 

\begin{figure}[!h]
\begin{center}
\begin{tabular}{c}
\includegraphics[height=3.75cm]{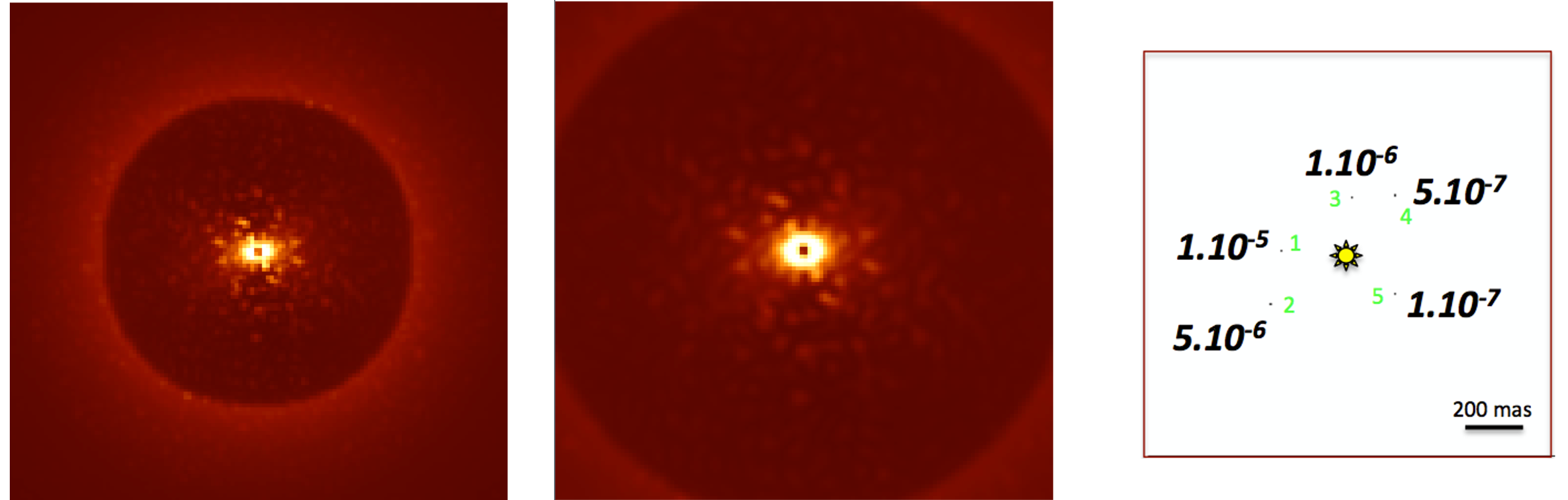}
\end{tabular}
\end{center}
\caption 
{ \label{fig:simuCI}
Simulated data for the inverse crime. Left: at the shortest wavelength ($950$ nm, Shannon sampled). Middle: at the longest wavelength ($1647.7$ nm). Right: Position, contrast and index of the five injected planetary signals.} 
\end{figure}

The static upstream and downstream phases, $\Phi_{up}$ and $\Phi_{do}$, are simulated by generating random phase screens following a power spectrum in $f^{-2}$ ($f$ being the spatial frequency) and of $30$ nm root mean square (rms) each at 950nm. 
As for the inversion, the multispectral cube is built from these aberrations by fixing the pupil diameter for all the wavelengths and adapting the focal plane sampling by zero padding the pupil image. 
In these simulations we built a six channels image cube ($\left[950,1083.6,1232.0,1365.6,1514.1,1647.7\right]$ nm) of infinitely small bandwidth. At these wavelengths, each image of the cube is defined with a spatial sampling proportional to an integer number of the Nyquist sampling. 
Photon noise is taken into account but no detector noise is added. 
Apart from the coronagraph and AO residuals, no other feature is taken into account (such as apodizer, central obstruction, spiders etc.). The simulated data are 6 images of $128 \times 128$ pixels with a scale of $12.25$ mas per pixel (see Fig.~\ref{fig:simuCI}-Left). Such simulations have also been used to validate the MEDUSAE concept in its previous publication\cite{Ygouf2013}. 

In the simulated data, we injected five planetary companions at various locations (below 400 mas) and various contrasts ranging from $10^{-5}$ to $10^{-7}$ (see Fig.~\ref{fig:simuCI}-Right).  

\subsection{Minimization strategy revisited}
The inversion strategy as previously published\cite{Ygouf2013} has been checked and improved. The improvement brought are mainly on the initialization procedure which is the most important part since it prevents from converging into one of the local minima instead of reaching the global minimum.

To summarize the previously published results\cite{Ygouf2013}, the initialization was performed along he following steps: (1) a very small random upstream phase screen (below 0.1 nm rms) is generated as a starting point, (2) the object map being set to zero (not estimated), the speckle field is estimated from the image at the shortest wavelength, (3) from the first upstream phase estimate, three equivalent phase maps (providing the exact same speckle field) are checked to obtain the one which minimizes better the criterion, (4) the steps (2) and (3) are repeated until convergence of the descent (user-defined value). After this initialization, the images at other wavelengths are incorporated one by one, and the minimization is ran including the object estimation (as described in Sec.~\ref{sect:medu1}).

In the new version, to help relaxing the first phase estimation, the step (2) is performed by alternatively estimating the object map, without spatial regularization nor support constraint. The step (3) is only performed once at the first estimation since it has been noticed that once the right phase is chosen, the descent carries on along the correct direction. These modified steps (initialization procedure) are performed using one then using two wavelengths (the shortest and the longest) to obtain a good estimate of the upstream phase before actually estimating the phase and the object alternatively. With this new initialization procedure, the step (4) is performed once, using all the available spectral channels all at once instead of incorporating them one by one. This brings a non-negligible gain in computation time (2h instead of 7h20 with 6 channels and images of $128\times 128$ pixels) and it appears that the estimation of the planet flux is more accurate (e.g. from 0.23\% instead of 0.59\% for the companion with a contrast of $10^{-5}$ and from a non-detected companion to 24\% for the companion at $10^{-7}$, both at a distance of 200 mas from the star). This non-negligible improvement is potentially due to the more accurate initialization procedure in combination with the increase of spectral diversity in one step to better constrain the estimations.

Using this improved initialization strategy, MEDUSAE provides the results presented in Tab.~\ref{tab:newCI}, in comparison with the former version of MEDUSAE\cite{Ygouf2013}. The relative mean square error (rMSE) of a parameter $X$ (in percentage) is defined as: 
\begin{equation}
\label{eq:rmse}
rMSE_X = \frac{ \sqrt{ \sum_\alpha [X(\alpha)-\hat{X}(\alpha)]^2}}{\sqrt{\sum_\alpha X(\alpha)^2} }  \times 100 \, ,
\end{equation}
with $X$ the true value used to simulate the data, and $\hat{X}$ its estimation by MEDUSAE. ${\phi}^{up}_{res}$ is the rms value of the difference between the true upstream phase and the estimated one.

\begin{table}[ht]
\caption{Results obtained at the end of the initialization procedure for the former and the new version of MEDUSAE. The convergence criterion has been set to $1.10^{-5}$.} 
\label{tab:newCI}
\begin{center}       
\begin{tabular}{|l | c| c| c| c| c| c| } 
\hline 
Version & Time [min] & Criterion $J_{data}$ &  $\mathrm{rMSE}_{fstar}$ & ${\phi}^{up}_{res}$ [rad] & $\mathrm{rMSE}_\Phi$ & $\mathrm{rMSE}_{SF}$ \\
\hline \hline 
New initialization: & 110 & $13,400$ & $0.14\%$   & $10^{-3}$   & $0.88\%$  & $0.15\%$\\
\hline 
Former initialization: & 35 & $268,213$ & $0.063\%$  & $0.03$      & $14.61\%$ & $0.46\%$\\
\hline 
\end{tabular}
\end{center}
\end{table} 

Note that the efficiency of this new initialization procedure is dependent upon the planetary signal intensity: if bright planetary companions are present in the data (with a contrast of about $3.10^{-4}$ and higher), it increases the final criterion (by about 5\%) and the rMSE of the estimated speckle field (by about 0.5\%). However, this effect remains small and can be bypassed by removing the bright signals before running MEDUSAE.

\subsection{Performance of MEDUSAE in the inverse crime framework}
\begin{figure}[!h]
\begin{center}
\begin{tabular}{c}
\includegraphics[height=7cm]{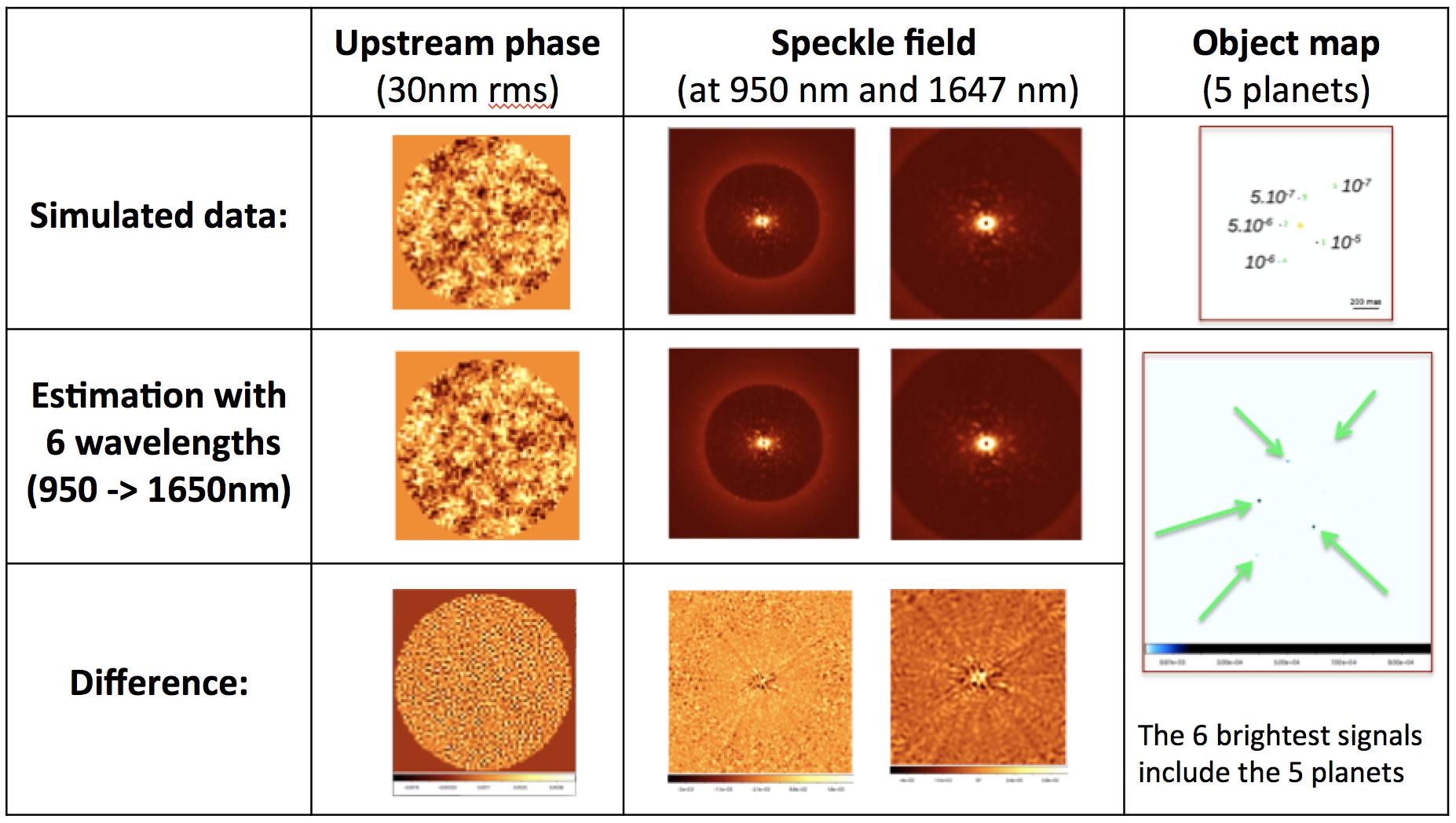}
\end{tabular}
\end{center}
\caption 
{ \label{fig:resCI}
Estimated upstream phase, speckle field and object map resulting from the MEDUSAE inversion applied to the simulated data described in Sec.~\ref{sect:medu2} and shown Fig.~\ref{fig:simuCI}.} 
\end{figure}

Applying MEDUSAE on the ideal simulated data, using the new inversion scheme, provides estimated speckle fields and object map shown on Fig.~\ref{fig:resCI}. The performance of the estimations are reported in Tab.~\ref{tab:resCI}\footnote{The results are provided, using IDL 8.5 on a 3.1 GHz Intel Core i5 macbook.}. The estimated upstream phase perfectly corresponds to the true one, thus, in the object map the brightest pixels indeed corresponds to a planetary signal. This object map can be thresholded by a $\sim 10^{-7}$ contrast which then provides only the planetary signals.

\begin{table}[ht]
\caption{Estimations provided by MEDUSAE applied on the simulated data shown in Fig.~\ref{fig:simuCI}, which are built using the ideal coronagraphic model (inverse crime). The convergence criterion has been set to $1.10^{-5}$.} 
\label{tab:resCI}
\begin{center}       
\begin{tabular}{|c|c|} 
\hline
Estimated parameter & Estimated value  \\
\hline\hline
Total inversion time & 2h \\
\hline
rMSE of the upstream phase & 0.74\%  \\
\hline
rMSE of the star spectrum & 0.01\% \\
\hline
rMSE of the multispectral speckle fields  & 0.10\% \\
\hline
Detected objects & all 5 companions (the 5 brightest signals) \\
\hline
rMSE of the contrasts of the 5 planets & (1): 0.23\%, (2): 0.11\%,(3): 1.46\%, (4): 16.5\%, (5) 24\% \\
\hline
\end{tabular}
\end{center}
\end{table} 

The typical sensitivity we can reach in this inverse crime framework is a contrast of $1.10^{-7}$ at 200 mas, with a precision of about 24\% on the contrast estimation and of one pixel on the position estimation. 

We use this new minimization strategy and the ideal simulated data as a baseline for the following sensitivity tests.

\subsection{Sensitivity to the input residual phase structure function, $D_{\Phi}$}
As previously explained, the turbulence residuals structure function map $D_{\Phi}$ must be given as an input to MEDUSAE. This parameters depends on the observation conditions and currently, it is not calibrated during the observation. To assess the sensitivity of MEDUSAE to this input, we ran MEDUSAE on various cases: the turbulence residuals structure function is null and the turbulence residuals structure function is generated with various wind and seeing characteristics.

To summarize our results, the turbulence residuals structure function is an extremely important input that must be carefully calibrated or computed beforehand. If it is not properly given in input, it completely affects the estimated speckle field (rMSE above 10\% and the final criterion is 700 to 5000 times higher), which makes it impossible to detect the planetary signals in the estimated object map: some light coming from the true residual phase, is not accounted for by the coronagraphic PSF model, and is therefore put in the estimated object map, thus hiding the true planetary companions (whatever the regularization applied to the object). Only companions whose contrast is already above the speckle floor can be detected but their flux estimation is biased ($>20\%$).

\subsection{Sensitivity to the input downstream phase, $\Phi_{do}$}
As previously explained, the downstream aberration map $\Phi_{do}$ must also be given as an input to MEDUSAE. The downstream phase should be relatively static, at least over one observation night, and it could be calibrated during the day. However the calibration might not be perfect or the aberrations might evolve between the calibration and the actual observation. To assess the impact of a mis-calibration of the downstream phase, we ran MEDUSAE on six different cases: with a null input map, with the true map with an error of 1\%, 10\%, 30\% or 50\% (obtained by adding a random normally distributed noise whose rms distribution is set so as to provide the error), and finally with a random phase screen having the same rms level and power spectral distribution.

To summarize our results, the downstream aberration map given in input to MEDUSAE must have the same spatial structure as the true one. If not, as for the turbulence residuals structure function, the light diffracted by the downstream aberrations is not taken into account by the model and thus is not part of the speckle field but is put inside the object map, thus preventing the detection and characterization of faint companions. As a result, the downstream phase must be accurately estimated beforehand. However, once its correct spatial distribution is found, a relative error on this map does not really affect the performance of MEDUSAE, as long as this error is below 50\%. 

These results are in line with preliminary results performed in the former version of MEDUSAE\cite{YgoufPhD}. 

\section{Application on realistic simulated data}
\label{sect:medu3}
\begin{figure}[!h]
\begin{center}
\begin{tabular}{c}
\includegraphics[height=3.5cm]{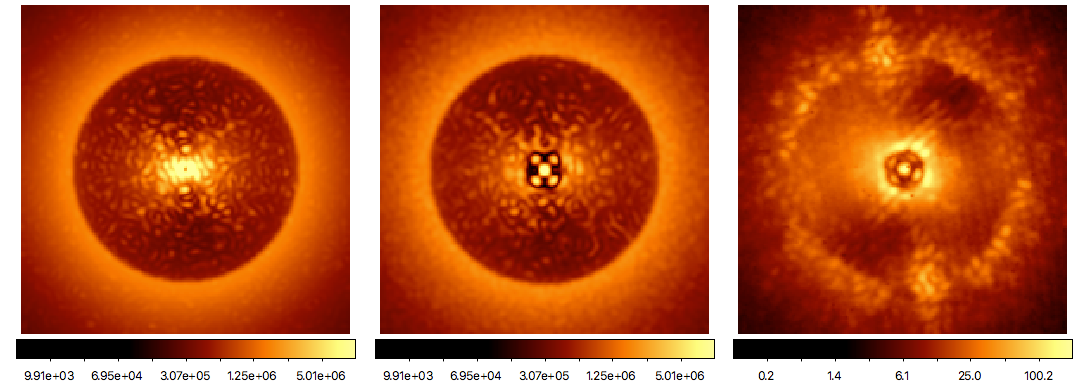}
\end{tabular}
\end{center}
\caption 
{ \label{fig:simureal}
Difference between the simulations and real data at $966$ nm. Simulated coronagraphic PSF obtained with the same aberrations, using the analytical expression of an ideal coronagraph\cite{Sauvage2010} (left) or using realistic simulations as described above (middle). For comparison, a typical SPHERE-IFS image of the star 51 Eridani\cite{Samland2017} is presented (right).} 
\end{figure}

When using an inverse problem approach, one has to be extremely cautious about the actual agreement between the model and the real data. As an example, the ANDROMEDA method\cite{Mugnier2009} suffered from mismatch between the model of noise and the real noise properties in the data. In the ANDROMEDA case, it is possible to catch up with this mismatch by applying spatial filtering to the images\cite{Cantalloube2015}, thus solving the mismatch in an empirical way. In MEDUSAE, considering the sensitivity to the $D_{\Phi}$ and $\Phi_{do}$, we decided to focus on the effect of a potential coronagraphic PSF model mismatch. To do so, we simulated data (so that both the $D_{\Phi}$ and the $\Phi_{do}$ are known), as close as possible to real data provided by the SPHERE-IFS instrument (Fig.~\ref{fig:simureal}-Right) which are significantly different from the coronagraphic PSF used in MEDUSAE's model (Fig.~\ref{fig:simureal}-Left). 

To simulate these so-called realistic multispectral data, we made the sum of $600$ short exposures coronagraphic PSF obtained, once again by propagating the light plane to plane in the Fraunhofer framework, as in Fig.~\ref{fig:SFmodel}. This time, we however took into account the diffractive effect of the focal plane mask coronagraph. In order to sample correctly the effect of the focal plane mask, we used the Matrix Fourier Transform\cite{Soummer2007} (MFT). The short exposure simulation takes as an input an AO residual phase screen, and not the structure function of the residual phase. For each short exposure, we simulated a random AO residual phase using the same method as in Sec.~\ref{sect:medu2}.

The entrance pupil is circular, and includes the central obstruction and the four spiders of the VLT as seen by SPHERE. The type of coronagraph simulated is the apodized Lyot coronagraph constituted of a pupil apodizer, an occulting focal plane mask of 185 mas diameter and a Lyot stop, as used in SPHERE. We therefore added the possibility to take into account exotic upstream and downstream pupils in the inversion procedure (now included in the model of the ideal coronagraphic image and its gradient).

To build the multispectral cube, contrary to the previous simulated data, we fixed the size of the images and changed the size of the pupils, accordingly to the wavelength. The static aberrations $\Phi_{up}$ and $\Phi_{do}$ are created as previously (Sec.~\ref{sect:medu2}). We also modified MEDUSAE to include the possibility of using images with any possible pixel sampling, and not only a multiple of the Nyquist sampling for all the channels.

The final image cube is constituted of $256\times256$ pixels images with 39 channels following the Y-H band (from 966 nm to 1640.7 nm) and spatial scaling (7.46 mas/px) of SPHERE-IFS. By comparing this simulated image (Fig.~\ref{fig:simureal}-Middle) with a real image from SPHERE-IFS (Fig.~\ref{fig:simureal}-Right), we see that the spatial structures\footnote{A specific feature not reproduced in our simulated images is the extra intensity of speckles in the $\sim -10\deg$ direction from vertical. This corresponds to the direction in the IFU plane which is aligned with the direction of the DM square actuator grid.} of the two images are similar, even in the central area (below $3\lambda/D$) where the coronagraph has a strong impact. We also checked that the spectral behavior is the same in the realistic simulations and real images from SPHERE-IFS.

The same planetary signals as on Fig.~\ref{fig:simuCI}-Right are injected in these realistic simulated images.

\subsection{Results from the MEDUSAE inversion}
We applied MEDUSAE on these realistic simulated images, using 13 spectral channels (every three other channel of SPHERE in the YH bands). During the inversion, the upstream phase is estimated so as to mainly reproduce the bright central part of the image which is however quite different from the model (see Fig.~\ref{fig:simureal}). As a consequence, the estimated upstream phase shows intense and discontinuous structures of 12 rad rms (instead of 0.20 rad rms for the true phase). In the focal plane this results into dark speckle-like structures all over the field of view. The object map contains spurious light from the ill-fitted speckle field, which makes it impossible to detect the planetary signals.

\subsection{Tuning the weight map}
The weight map is the inverse of the noise variance, $1/\sigma_n^2$, and is a multiplicative factor in the criterion expression in Eq.~(\ref{eq:criterion}), dependent upon both the spectral and spatial dimension. Thus, we can force the inversion to be less or more stringent at some location and at some wavelength by tuning this weight map. 

Because the main visible difference between the realistic data (Fig.~\ref{fig:model}-Middle) and our model (Fig.~\ref{fig:model}-Left) is at the center of the image, where the coronagraph has a strong impact, we decided to set the weight map to zero below a separation of $3\lambda/D$.

\begin{figure}[!h]
\begin{center}
\begin{tabular}{c}
\includegraphics[height=7cm]{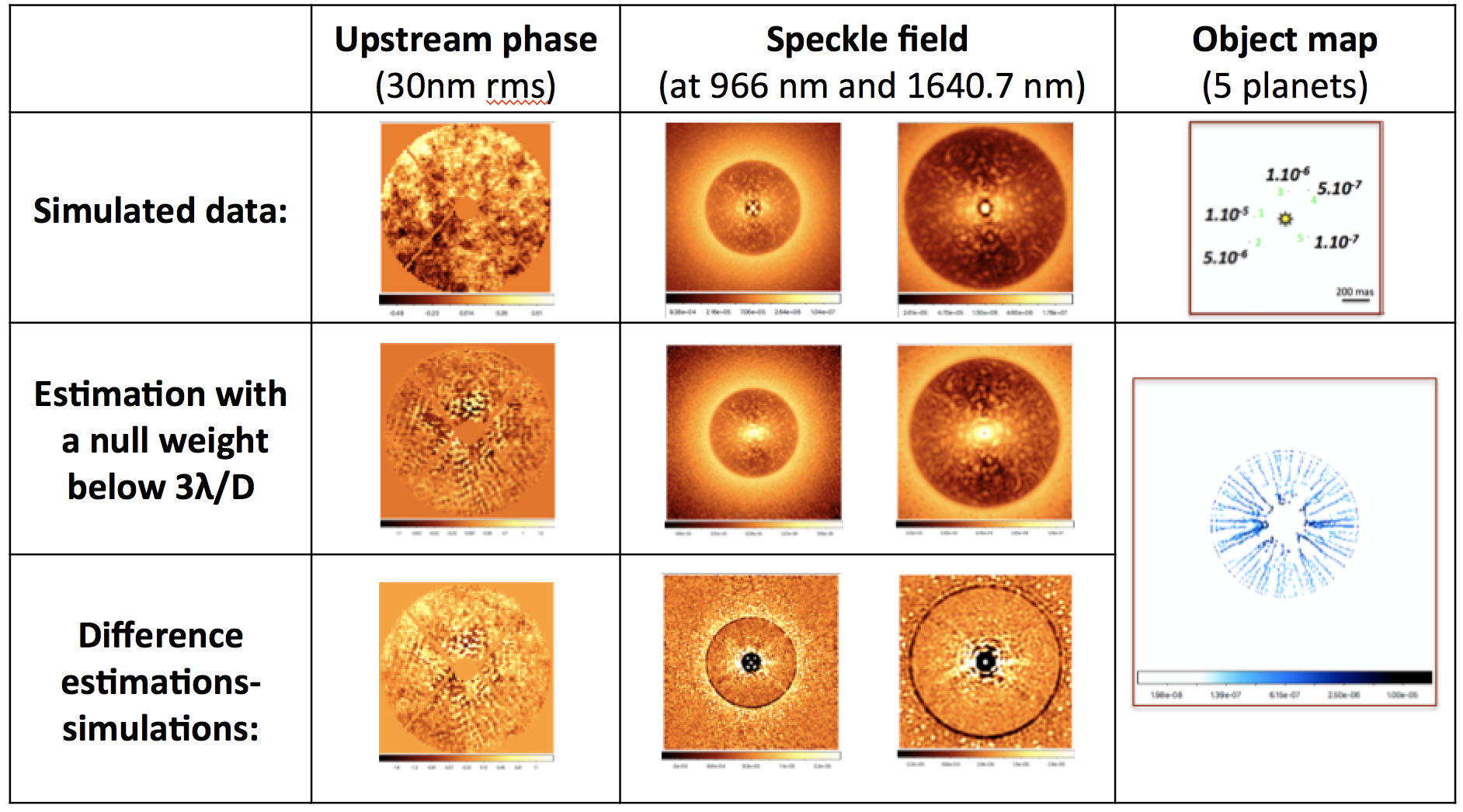}
\end{tabular}
\end{center}
\caption 
{ \label{fig:resRS}
Results from the MEDUSAE inversion applied to the realistic simulated data cube as on Fig.~\ref{fig:simureal}-Middle.} 
\end{figure}

In that case, the estimated upstream phase remains quite different from the true one but the resulting speckle field is closer to the true one, as shown on Fig.~\ref{fig:resRS}. The performance of the inversion are gathered in Tab.~\ref{tab:resRS}. Even though the speckle field is estimated with a 18\% accuracy, too much residual starlight is put in the object map (which consequently shows radial extensions, as visible on Fig.~\ref{fig:resRS}) and it is impossible to detect the presence of faint planetary signals among these residuals.   

\begin{table}[ht]
\caption{Estimations from MEDUSAE applied on the realistic simulated data shown in Fig.~\ref{fig:simureal}-Middle, with a weight map set to zero in the central region ($< 3\lambda/D$).  For comparison, the middle column presents the results obtained on the same data but using the ideal coronagraphic model instead of the APLC (Fig.~\ref{fig:simureal}-Left). In both cases, the image cube consists of 13 channels of $256\times 256$ pixels and the convergence criterion has been set to $1.10^{-5}$.}
\label{tab:resRS}
\begin{center}       
\begin{tabular}{|c|c|c|} 
\hline
Estimated parameter & Ideal coronagraph simulations  & Realistic APLC simulations \\
\hline\hline
Total inversion time & 5h30 & 8h \\
\hline
rMSE of the upstream phase & $1\%$ & $> 100 \%$  \\
\hline
rMSE of the star spectrum & $0.01\%$ & $66\%$ \\
\hline
rMSE of the multispectral speckle fields & $0.01\%$ & $18\%$ \\
\hline
Detected objects & all (5 brightest) & none \\
\hline
rMSE of the contrasts of the 5 planets & 2.5\%, 13\%, 3\%, 4\%, 26\% & N/A \\
\hline
\end{tabular}
\end{center}
\end{table} 


\section{Conclusion and perspectives}
MEDUSAE is an image processing technique dedicated to multispectral images which takes into account the knowledge about the instrument design. Instead of regarding the speckle as noise to be removed, MEDUSAE carefully estimate those, jointly with the object map, in a Bayesian framework. Developing such algorithm is important to match recent and future high contrast instrument images and increase their sensitivity. Moreover, as MEDUSAE is a passive technique, it can be used on any data, as long as observation strategy and dedicated calibrations are taken into account.

In this communication, we showed that although this method proves efficient on simulations, the model used for the inversion does need to be refined in order to better match the real data. We are currently implementing a new analytical long exposure PSF model, published in Ref.~\citenum{Herscovici2017}, that has been successfully implemented in the similar algorithm COFFEE\cite{Paul2013}. After completing the implementation of the realistic coronagraph PSF model, we will study how to reliably obtain the residual phase structure function and downstream phase map.

\appendix   

\section{Inside MEDUSAE, the inversion procedure in practice}
\label{sect:annexe}
In this annexe, we summarize how the minimization takes place in the MEDUSAE algorithm.

Estimating the speckle field by multispectral phase retrieval:\\
Assuming that the object map is known, the descent of the criterion $J$ is made via an iterative local descent algorithm which uses the analytical expression of the gradient of the criterion. The gradient of the PSF to the estimated phase is analytically computed from the analytical expression of the long exposure PSF\cite{Ygouf2013}. 
Inside MEDUSAE, we use the Variable Metric with Limited Memory and Bounds (VMLM-B\cite{Thiebaut2002}), which has the advantage of incorporating bounds on every parameter to better constrain the descent. This quasi-Newton method has for main advantage to reach the minimum quite fast given the numerous data point processed by MEDUSAE. Its main drawback is that it is sensitive to the numerous local minima and thus to the starting point of the descent.\\
The star spectrum is estimated jointly with a specific regularization at each wavelength, in order to avoid the divergence of the estimated flux.

Estimating the object map via a non-myopic deconvolution:\\
Assuming that the speckle field is known, the object is estimated via a multispectral non-myopic deconvolution (the criterion is convex, providing a unique solution). The spatial regularization applied to the object estimation has to be chosen accordingly to the object sought. The implemented regularization in MEDUSAE is a L1-L2 regularization which preserves the edges while smoothing the noise based on the gradient intensity in the image. The hyper-parameters can be tuned to increase the regularization toward a L2 (smoothing small pixel to pixel intensity variations) or toward a L1 (preserving large variations) norm. A support constraint has also been added to consider that all the light which is outside of the AO correction zone belongs to the star. The latter constraint helps estimating the speckle field which depends on both the star spectrum and the upstream phase.



\bibliography{report}   

\begin{thebibliography}{10}

\bibitem{Cheetham2016}
A.~C. {Cheetham}, J.~{Girard}, S.~{Lacour}, {\em et~al.}, ``{Sparse aperture
  masking with SPHERE},'' in {\em Optical and Infrared Interferometry and
  Imaging V},  {\em Proc. SPIE} {\bf 9907}, 99072T  (2016).

\bibitem{Samland2017}
M.~{Samland}, P.~{Molli{\`e}re}, M.~{Bonnefoy}, {\em et~al.}, ``{Spectral and
  atmospheric characterization of 51 Eridani b using VLT/SPHERE},'' {\em A\&A}
  {\bf 603}, A57  (2017).

\bibitem{Antichi2008}
J.~Antichi, R.~G. Gratton, R.~U. Claudi, {\em et~al.}, ``Sphere ifs optical
  concept description and design overview,''  (2008).

\bibitem{Macintosh2008}
B.~A. {Macintosh}, J.~R. {Graham}, D.~W. {Palmer}, {\em et~al.}, ``{The Gemini
  Planet Imager: from science to design to construction},'' in {\em Adaptive
  Optics Systems},  {\em Proc. SPIE} {\bf 7015}, 701518  (2008).

\bibitem{Peters-Limbach2013}
M.~A. {Peters-Limbach}, T.~D. {Groff}, N.~J. {Kasdin}, {\em et~al.}, ``{The
  optical design of CHARIS: an exoplanet IFS for the Subaru telescope},'' in
  {\em Techniques and Instrumentation for Detection of Exoplanets VI},  {\em
  Proc. SPIE} {\bf 8864}, 88641N  (2013).

\bibitem{Hinkley2007}
S.~{Hinkley}, B.~R. {Oppenheimer}, R.~{Soummer}, {\em et~al.}, ``{Temporal
  Evolution of Coronagraphic Dynamic Range and Constraints on Companions to
  Vega},'' {\em ApJ} {\bf 654}, 633--640  (2007).

\bibitem{Beuzit2008}
J.-L. {Beuzit}, M.~{Feldt}, K.~{Dohlen}, {\em et~al.}, ``{SPHERE: a 'Planet
  Finder' instrument for the VLT},'' in {\em Ground-based and Airborne
  Instrumentation for Astronomy II},  {\em Proc. SPIE} {\bf 7014}, 701418
  (2008).

\bibitem{Racine1999}
R.~{Racine}, G.~A.~H. {Walker}, D.~{Nadeau}, {\em et~al.}, ``{Speckle Noise and
  the Detection of Faint Companions},'' {\em Pub. Astron. Soc. Pacific.} {\bf
  111}, 587--594  (1999).

\bibitem{Sparks2002}
W.~B. {Sparks} and H.~C. {Ford}, ``{Imaging Spectroscopy for Extrasolar Planet
  Detection},'' {\em ApJ} {\bf 578}, 543--564  (2002).

\bibitem{Sauvage2010}
J.-F. {Sauvage}, L.~M. {Mugnier}, G.~{Rousset}, {\em et~al.}, ``{Analytical
  expression of long-exposure adaptive-optics-corrected coronagraphic image
  First application to exoplanet detection},'' {\em Journal of the Optical
  Society of America A} {\bf 27}, A157  (2010).

\bibitem{Mugnier2004}
L.~M. {Mugnier}, T.~{Fusco}, and J.-M. {Conan}, ``{MISTRAL: a myopic
  edge-preserving image restoration method, with application to astronomical
  adaptive-optics-corrected long-exposure images},'' {\em Journal of the
  Optical Society of America A} {\bf 21}, 1841--1854  (2004).

\bibitem{Ygouf2013}
M.~{Ygouf}, L.~M. {Mugnier}, D.~{Mouillet}, {\em et~al.}, ``{Simultaneous
  exoplanet detection and instrument aberration retrieval in multispectral
  coronagraphic imaging},'' {\em A\&A} {\bf 551}, A138  (2013).

\bibitem{CantalloubePhD}
F.~Cantalloube, {\em {Detection and characterization of exoplanets in high
  contrast images by the inverse problem approach}}.
\newblock Theses, {Universit{\'e} Grenoble Alpes}  (2016).

\bibitem{Devaney2017}
N.~{Devaney} and {\'E}.~{Thi{\'e}baut}, ``{PeX 1. Multi-spectral expansion of
  residual speckles for planet detection},'' {\em ArXiv e-prints}   (2017).

\bibitem{Perrin2003}
M.~D. {Perrin}, A.~{Sivaramakrishnan}, R.~B. {Makidon}, {\em et~al.}, ``{The
  Structure of High Strehl Ratio Point-Spread Functions},'' {\em ApJ} {\bf
  596}, 702--712  (2003).

\bibitem{Jolissaint2006}
L.~{Jolissaint}, J.-P. {V{\'e}ran}, and R.~{Conan}, ``{Analytical modeling of
  adaptive optics: foundations of the phase spatial power spectrum approach},''
  {\em Journal of the Optical Society of America A} {\bf 23}, 382--394  (2006).

\bibitem{YgoufPhD}
M.~Ygouf, {\em {New method of multispectral image post-processing based on an
  instrument model for high contrast imaging systems : Application to exoplanet
  detection}}.
\newblock Theses, {Universit{\'e} de Grenoble}  (2012).

\bibitem{Mugnier2009}
L.~M. {Mugnier}, A.~{Cornia}, J.-F. {Sauvage}, {\em et~al.}, ``{Optimal method
  for exoplanet detection by angular differential imaging},'' {\em Journal of
  the Optical Society of America A} {\bf 26}, 1326  (2009).

\bibitem{Cantalloube2015}
F.~{Cantalloube}, D.~{Mouillet}, L.~M. {Mugnier}, {\em et~al.}, ``{Direct
  exoplanet detection and characterization using the ANDROMEDA method:
  Performance on VLT/NaCo data},'' {\em A\&A} {\bf 582}, A89  (2015).

\bibitem{Soummer2007}
R.~{Soummer}, L.~{Pueyo}, A.~{Sivaramakrishnan}, {\em et~al.}, ``{Fast
  computation of Lyot-style coronagraph propagation},'' {\em Optics Express}
  {\bf 15}, 15935  (2007).

\bibitem{Herscovici2017}
O.~{Herscovici-Schiller}, L.~M. {Mugnier}, and J.-F. {Sauvage}, ``{An analytic
  expression for coronagraphic imaging through turbulence. Application to
  on-sky coronagraphic phase diversity},'' {\em Mon. Not. R. astr. Soc.} {\bf
  467}, L105--L109  (2017).

\bibitem{Paul2013}
B.~{Paul}, L.~M. {Mugnier}, J.-F. {Sauvage}, {\em et~al.}, ``{High-order myopic
  coronagraphic phase diversity (COFFEE) for wave-front control in
  high-contrast imaging systems},'' {\em Optics Express} {\bf 21}, 31751
  (2013).

\bibitem{Thiebaut2002}
E.~{Thiebaut}, ``{Optimization issues in blind deconvolution algorithms},'' in
  {\em Astronomical Data Analysis II},  J.-L. {Starck} and F.~D. {Murtagh},
  Eds., {\em Proc. SPIE} {\bf 4847}, 174--183  (2002).

\end{thebibliography}
\bibliographystyle{spiejour}   


\end{document}